\documentclass[lettersize,journal]{IEEEtran}

\usepackage{amsmath,amsfonts,amssymb}
\usepackage{amsthm}
\theoremstyle{plain}
\newtheorem{theorem}{Theorem} 
\newenvironment{customproof}[1][Proof]{\noindent\textbf{#1.} }{\hfill$\square$\par}

\usepackage{graphicx}
\usepackage{tikz}
\usepackage{subcaption}
\usepackage{pgfplots}
\usepackage{pgfplotstable}
\usepgfplotslibrary{groupplots}
\usetikzlibrary{matrix}

\usepackage{array}
\usepackage{multirow}
\usepackage{booktabs}
\usepackage{makecell}
\usepackage{adjustbox}

\usepackage[linesnumbered,ruled,vlined]{algorithm2e}

\usepackage{caption}
\usepackage[caption=false,font=normalsize,labelfont=sf,textfont=sf]{subfig}

\usepackage{textcomp}
\usepackage{soul}
\usepackage{color}
\usepackage{enumitem}

\usepackage{cite}
\usepackage{url}

\usepackage{stfloats}
\usepackage{verbatim}
\usepackage{wrapfig}
\usepackage{bbding}

\definecolor{1}{RGB}{141,211,199}
\definecolor{2}{RGB}{255,255,179}
\definecolor{3}{RGB}{190,186,218}
\definecolor{44}{RGB}{161,63,44}
\definecolor{5}{RGB}{156,209,255}
\definecolor{66}{RGB}{253,180,98}
\definecolor{77}{RGB}{179,222,105}
\definecolor{8}{RGB}{251,154,153}
\definecolor{99}{RGB}{31,120,180}
\definecolor{6}{RGB}{107,94,139}
\definecolor{9}{RGB}{78,129,121}
\definecolor{7}{RGB}{205,135,113}
\definecolor{4}{RGB}{127,66,82}
\definecolor{orange}{RGB}{43, 112, 47}
\definecolor{black}{RGB}{43,51,123}

\begin{document}

\title{Context-Adaptive Graph Neural Networks for Next POI Recommendation}

\author{Yu Lei, Limin Shen, Zhu Sun$^*$, Tiantian He$^*$, Yew-Soon Ong
\thanks{*Corresponding authors.}
\IEEEcompsocitemizethanks{
\IEEEcompsocthanksitem 
Yu Lei and Limin Shen are with the School of Information Science and Engineering, Yanshan University, and Key Lab for Software Engineering of Hebei Province, China. E-mail: leiyu1111@stumail.ysu.edu.cn, shenllmm@sina.com.
\IEEEcompsocthanksitem 
Zhu Sun is with Pillar of Information Systems Technology and Design, Singapore University of Technology and Design, Singapore. E-mail: sunzhuntu@gmail.com.
\IEEEcompsocthanksitem 
Tiantian He is with the Centre for Frontier AI Research, Institute of High Performance Computing, and Singapore Institute of Manufacturing Technology, A*STAR, Singapore. E-mail: he\_tiantian@cfar.a-star.edu.sg.
\IEEEcompsocthanksitem 
Yew-Soon Ong is with the Institute of High Performance Computing and Centre for Frontier AI Research, A*STAR, and also with the College of Computing and Data Science, Nanyang Technological University, Singapore. E-mail: asysong@ntu.edu.sg.
}
}%



\maketitle

\begin{abstract}
Next Point-of-Interest (POI) recommendation is a critical task in location-based services, aiming to predict users' next visits based on their check-in histories. While many existing methods leverage Graph Neural Networks (GNNs) to incorporate collaborative information and improve recommendation accuracy, most of them model each type of context using separate graphs, treating different factors in isolation. This limits their ability to model the co-influence of multiple contextual factors on user transitions during message propagation, resulting in suboptimal attention weights and recommendation performance.
Furthermore, they often prioritize sequential components as the primary predictor, potentially undermining the semantic and structural information encoded in the POI embeddings learned by GNNs.
To address these limitations, we propose a \underline{C}ontext-\underline{A}daptive \underline{G}raph \underline{N}eural \underline{N}etworks (CAGNN) for next POI recommendation, which 
dynamically adjusts attention weights using edge-specific contextual factors and enables mutual enhancement between graph-based and sequential components.
Specifically, CAGNN introduces (1) a \textit{context-adaptive attention} mechanism that jointly incorporates different types of contextual factors into the attention computation during graph propagation, enabling the model to dynamically capture collaborative and context-dependent transition patterns; (2) a \textit{graph-sequential mutual enhancement} module, which aligns the outputs of the graph- and sequential-based modules via the KL divergence, enabling mutual enhancement of both components. 
Experimental results on three real-world datasets demonstrate that CAGNN consistently outperforms state-of-the-art methods.
Meanwhile, theoretical guarantees are provided that our context-adaptive attention mechanism improves the expressiveness of POI representations.
\end{abstract}


\section{Introduction}

Next Point-of-Interest (POI) recommendation has become an essential service in location-based applications~\cite{ye2024adaptive, rao2025next, chen2025next}, aiming to predict a user’s next visit based on historical trajectories by analyzing their movement patterns and contextual preferences. 
Traditional methods primarily rely on sequential-based models, such as Markov Chains (MC)~\cite{rendle2010factorizing}, recurrent neural networks (RNNs)~\cite{feng2018deepmove,huang2019attention}, and Transformers~\cite{zhang2022next,luo2021stan}, to capture temporal dependencies and transition patterns in user check-in sequences\footnote{In our study, the two terms `trajectory' and `sequence' are exchangeable.}. 
While these models effectively learn user behaviors within individual sequences, they struggle to incorporate collaborative signals when learning POI representations.
%
To address these limitations, recent studies{~\cite{li2021discovering,lai2024disentangled,sun2023multi}} have explored Graph Neural Network (GNN)-based approaches for next POI recommendation. 
These methods construct POI relation graphs from user check-in data to capture collaborative patterns among all users. 
GNNs then learn structural and relational information within these graphs to generate POI embeddings\footnote{{The two terms ‘embedding' and ‘representation' are also exchangeable.}}. These embeddings are subsequently integrated into sequential models (e.g., Long Short-Term Memory (LSTM) or Transformer) to capture users' sequential preferences. 
Despite their effectiveness, existing GNN-based next POI recommendation methods face two major limitations.

\begin{figure*}[t]
\centering
\includegraphics[width=0.8\linewidth]{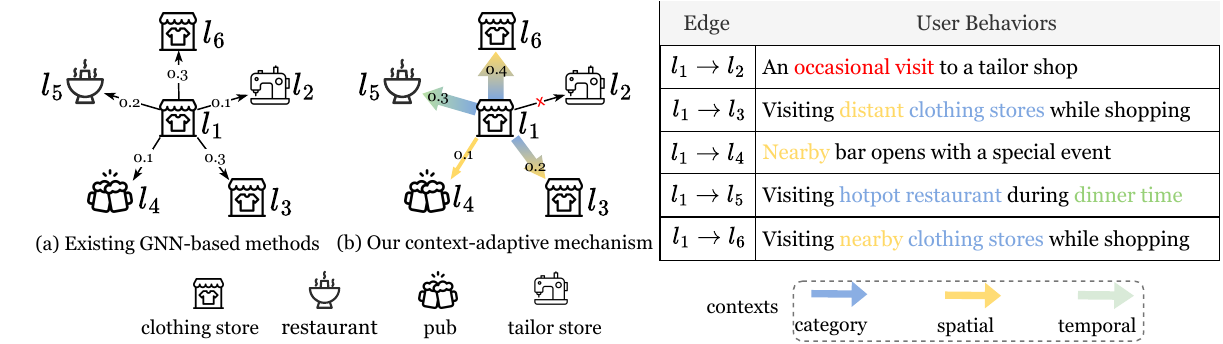}
\caption{A running example of a POI-POI transition graph: (a) with attention weights based on node embedding similarity and (b) with context-adaptive attention weights. Arrow colors represent different contextual factors influencing the transitions, while arrow widths indicate the importance of each connected node to the central node, with wider arrows signifying greater importance. Mixed-color arrows indicate influence from multiple contextual factors. ‘\textcolor{red}{$\times$}' denotes the filtered edge.}
\label{fig:running_example}
\vspace{-0.1in}
\end{figure*}

First, \textbf{lack of context-aware edge modeling.}
GNN-based next POI recommendation methods commonly adopt graph attention networks (GATs) or graph convolutional networks (GCNs) to capture collaborative signals among POIs~\cite{yang2022getnext, lim2020stp}. 
However, standard GATs are node-centric, computing attention weights based only on node embeddings while ignoring edge-specific contextual factors such as spatial distance, time, and POI category~\cite{li2021discovering,wang2024bi}.
Although some GCN-based methods incorporate contextual factors, they typically model only a single type of context per graph, with each graph constructed using predefined rules based on that specific context, such as transition frequency~\cite{yang2024siamese,yang2022getnext}, spatial distance~\cite{qin2023diffusion,wang2022graph}, or temporal intervals~\cite{lim2020stp,wang2023global}.
Consequently, these approaches cannot comprehensively model the co-influence of multiple types of contextual factors on user transitions during message propagation.

In practice, this limitation becomes more significant since user transitions are usually shaped by multiple, interrelated contextual factors. 
As illustrated in Fig.~\ref{fig:running_example}(a), the transition from a clothing store $l_1$ to a hotpot restaurant $l_5$ is likely driven by both time (e.g., dinner time) and category (e.g., restaurant). Similarly, a transition from $l_1$ to a nearby clothing store $l_3$ may be influenced by both distance and category. However, GNN-based methods often overlook the co-influence of interrelated contextual factors and tend to treat all transitions uniformly, failing to distinguish their varying degrees of importance.
For instance, GAT may assign a moderate attention weight (e.g., 0.1) to an infrequent transition from $l_1$ to a tailor shop $l_2$. Similarly, for two clothing stores $l_3$ and $l_6$ connected to $l_1$, GAT may assign similar attention weights (e.g., 0.3) due to embedding similarity, even though $l_6$ is geographically distant and less likely to be the immediate next destination compared to the nearby $l_3$. 
In contrast, our proposed context-adaptive attention mechanism jointly models node features and multiple types of contextual factors. This enables the model to assign more accurate attention weights by emphasizing contextually relevant transitions and suppressing less meaningful ones, as shown in Fig.~\ref{fig:running_example}(b), thereby effectively capturing user behavior semantics for more accurate recommendations.

Second, \textbf{dominated sequential models}. Under the GNN-based frameworks, they treat the sequential models (e.g., LSTM, Transformer) as the primary predictor, using graph-based POI embeddings only as auxiliary input. This design causes the sequential component to dominate the prediction, potentially undermining the semantic and structural information encoded in the POI embeddings learned by GNNs. Thus, a seamless integration of the graph and sequential components is necessary to fully exploit their complementary strengths for more effective recommendations.


To address the above limitations, we propose a \underline{C}ontext-\underline{A}daptive \underline{G}raph \underline{N}eural \underline{N}etworks (CAGNN) for next POI recommendation, which jointly incorporates multiple types of edge-specific contextual factors to dynamically adjust attention weights and enables mutual enhancement between graph-based and sequential components, thereby improving recommendation accuracy.
It consists of three key modules:
(1) \textit{Graph-based User Preference Extractor with Context-Adaptive Attention}, which dynamically adjusts standard GAT attention weights by jointly integrating category, spatial, and temporal contextual factors. 
This reinforces contextually relevant transitions and suppresses less informative ones, thereby enhancing the modeling of collaborative and context-dependent patterns and improving prediction accuracy;
(2) \textit{Sequential User Preference Extractor}, which leverages a Transformer layer to capture each user's sequential preference based on their short-term check-in history; 
and (3) \textit{Graph-Sequential Mutual Enhancement Module}, which aligns POI embeddings from both the graph- and sequential-based extractors using the KL divergence, enabling the graph component to incorporate sequential patterns and the sequential component to integrate collaborative signals for mutual enhancement.

In summary, our main contributions lie four-fold. 
\begin{itemize}
\item We are the first to identify two key limitations of existing GNN-based next POI recommendation models, including the lack of context-aware edge modeling and dominated sequential models.
\item We propose CAGNN, a novel framework that not only dynamically adjusts graph attention weights based on varying contextual factors, but also enables mutual enhancement between graph and sequential representations to better model user dynamics and collaborative behavior.
\item We perform a theoretical analysis to prove that our context-adaptive attention mechanism enhances the expressiveness of POI representations, thereby improving the model's recommendation accuracy.
\item We conduct extensive experiments on three real-world datasets, demonstrating that our proposed CAGNN significantly outperforms state-of-the-art methods (SOTAs), achieving average improvements of 10.47\% and 14.85\% on HR and NDCG, respectively. 
\end{itemize}
\section{Related Works}

\subsection{Next POI Recommednation}
Next Point-of-Interest (POI) recommendation aims to predict a user's next destination by analyzing their mobility trajectories and contextual factors. To capture sequential patterns, various models have been explored, ranging from classical probabilistic approaches to advanced deep learning architectures.
MC-based models~\cite{rendle2010factorizing,cheng2013you} were among the earliest approaches. They rely on transition probabilities derived from historical check-in data to predict a user’s next POI based on past locations\footnote{In our study, the two terms `POI' and `location' are exchangeable.}.
RNNs and their variants, such as LSTM and Gated Recurrent Units (GRUs), were later introduced to model sequential dependencies in user mobility~\cite{liu2016predicting,feng2018deepmove,huang2019attention,zhang2020interactive,zhao2020where}. These models improved upon MC-based methods by capturing long-term dependencies in check-in sequences.
Self-attention and Transformer-based models~\cite{lin2021pre,zhang2022next,yan2023spatio,kang2018self,luo2021stan,yang2022getnext} have recently gained popularity due to their ability to effectively model both short- and long-range dependencies in sequential data, outperforming traditional RNN-based recommenders.

Beyond optimizing sequential modeling, research has focused on integrating contextual factors such as spatial features (e.g., proximity)~\cite{feng2017poi2vec,liu2019geo,lian2020geography, sun2024city}, temporal patterns (e.g., periodicity)~\cite{zhao2016stellar,yu2020category,feng2024rotan}, POI categories~\cite{liao2018predicting,wu2020personalized,yao2017sermA}, and the joint effects of various contextual factors~\cite{cui2021st, sun2023multi,xie2024hierarchical} to enhance recommendation accuracy. Further advancing the field, researchers are now exploring novel paradigms by integrating techniques like Large Language Models (LLMs)~\cite{feng2024move, li2024large, wang2024secor}, meta-learning~\cite{cui2021sequential,wang2023meta}, and reinforcement learning~\cite{zhang2022conversation, du2023cross}.

\subsection{Next POI Recommendation with Graph Neural Networks}
GNNs have emerged as powerful tools for modeling complex interactions in recommender systems, particularly in capturing global user–item relationships~\cite{sun2020neighbor,rao2022graph,chen2025next,chen2024graph,huang2021position,zhu2025multi}. 
In the context of next POI recommendation, GNNs propagate information through POI-POI graph structures, enabling nodes to iteratively aggregate collaborative signals from their neighbors and learn complex connectivity patterns~\cite{wang2022graph, wang2022learning, liu2023mandari}.

Early works~\cite{yang2024siamese,cao2023improving,yang2022getnext} construct POI–POI graphs based on user transition behaviors. For example, GETNext~\cite{yang2022getnext} encodes transition information via graph convolution on customized trajectory graphs, while MCMG~\cite{sun2023multi} builds a POI transition graph to capture collaborative signals across all users.
Later, several studies~\cite{qin2023diffusion, wang2022learning, yin2023next} incorporate a single type of context into graph construction or modeling to better learn user preferences. For instance, SGRec~\cite{li2021discovering} enhances GAT with category information to learn category-aware transition patterns, and AGRAN~\cite{wang2023adaptive} incorporates spatial context by dynamically constructing POI-POI graphs based on geographical similarity among POIs.

To further improve the recommendation performance, more recent approaches~\cite{wang2022graph, wang2023global, xu2023revisiting, rao2025next} extend this line of work by integrating multiple types of contextual factors. These methods typically construct separate POI–POI graphs for each context and combine the learned representations for the final recommendation.
For example, STP-UDGAT~\cite{lim2020stp} integrates spatial, temporal, and frequency-based graphs to model user behavior patterns under different contextual factors. 
DCHL~\cite{lai2024disentangled} constructs multi-view disentangled hypergraphs based on views of collaborative, transitional, and geographical signals.
BiGSL~\cite{wang2024bi} introduces three bi-level graphs under the spatial, temporal, and transition feature views to hierarchically capture fine-to-coarse connections between POIs and prototypes.

Despite these advancements, most existing methods model each context on separate graphs, treating different contextual factors in isolation. 
This limits their ability to model the co-influence of multiple contextual factors on user transitions during message propagation, which leads to suboptimal attention weights and recommendation performance.
Moreover, many existing frameworks prioritize sequential components as the primary predictor, potentially undermining the semantic and structural information encoded in the POI embeddings learned by GNNs.
To address these limitations, we jointly incorporate multiple contextual factors, such as category, spatial, and temporal information, into the attention computation to capture the co-influence of different contextual factors on user transitions. Additionally, we design an integration mechanism that seamlessly combines graph-based and sequential components to mutually enhance the learning processes of both components.

\section{Method}

In this section, we introduce our proposed CAGNN for the next POI recommendation. 

\subsection{Preliminaries and Model Overview}

\begin{table}
\centering
\caption{Important notations and descriptions.}
\addtolength{\tabcolsep}{-2pt}
\begin{tabular}{l|l}
\specialrule{.15em}{.15em}{.15em}
\textbf{Notations}  & \textbf{Descriptions}\\ 
\specialrule{.1em}{.1em}{.1em}
$u, v, t, c, d $ & \makecell[l]{user, POI, time slot, category of POI and\\ distance interval}\\
$\mathbf{u},\mathbf{v},\mathbf{t},\mathbf{c},\mathbf{d}\in \mathbb{R}^{dim}$  & \makecell[l]{embeddings of user, POI, time slot, category \\of POI and distance interval} \\
$\mathbf{V}, \mathbf{C}, \mathbf{T}, \mathbf{D}$ & \makecell[l]{embedding matrix of POI, POI category, \\time slot and distance interval}\\
$\mathcal{U}$, $\mathcal{V}$, $\mathcal{C}$, $\mathcal{T}$, $\mathcal{D}$ & \makecell[l]{sets of users, POIs, categories, time slot and \\distance interval}\\
$\mathcal{G}$ & the POI-POI transition graph \\
$\mathcal{V},\mathcal{E}$ & sets of nodes and edges in $\mathcal{G}$\\
$\mathcal{S}^{u,i}= \{S_{t_1}^u,\cdots, S_{t_k}^u$\} & the $i$-th POI sequence of user $u$\\
$S_{t_k}^u = (u, v_{t_k}^u, t_{k})$ & \makecell[l]{the check-in record of POI $v$ performed by\\ user $u$ at time $t_k$}\\
$\mathbf{d}_{i,:}\in \mathbb{R}^{\operatorname{max}(\mathcal{D})}$ & \makecell[l]{the transition distribution for each distance \\interval when POI $v_i$ is the neighbor node}\\
$\mathbf{t}^{v_i} \in \mathbb{R}^{24}$ & \makecell[l]{the hourly check-in frequency of POI $v_i$ in\\ each hour of the day} \\
\specialrule{.15em}{.15em}{.15em}
\end{tabular}\label{table:notations}
\vspace{-0.1in}
\end{table}

\smallskip\noindent\textbf{Notations.} 
Let $\mathcal{U}=\{u_1,u_2,\cdots,u_{|\mathcal{U}|}\}$, $\mathcal{V}=\{v_1,v_2,\cdots,v_{|\mathcal{V}|}\}$, and $\mathcal{C}=\{c_1,c_2,\cdots,c_{|\mathcal{C}|}\}$, denote the sets of users, POIs, and categories of POIs, respectively.
We map one day into 24 slots $\mathcal{T}=\{t_1,t_2,\cdots,t_{24}\}$. 
A check-in record $(u,v,t)$ indicates user $u$ visited POI $v$ at time $t$. For each user, we follow SOTAs~\cite{zhang2020interactive, feng2024rotan} to chronologically order all her historical check-in records and then split them into sequences by day. 
The $i$-th check-in sequence of user $u$ is denoted as $\mathcal{S}^{u,i}$=$\{S_{t_1}^u, S_{t_2}^u,\cdots, S_{t_k}^u\}$, and $S_{t_k}^u$=$(u, v_{t_k}^u, t_{k})$. 
We construct a directed POI-POI transition graph $\mathcal{G}=\left(\mathcal{V}, \mathcal{E}\right)$ based on all users' POI check-in trajectories, where $\mathcal{V}$ and $\mathcal{E}$ are the sets of POIs and edges, respectively. An edge exists from POI $v_i$ to $v_j$ if $v_j$ is the next POI visited after $v_i$ by the same user $u$. 
Given the trajectory $\mathcal{S}^{u, i}$ of user $u$ and the graph $\mathcal{G}$, our goal is to recommend to user $u$ a list of top-ranked POIs at the next time slot $t_{k+1}$. 
Important notations are summarized in Table~\ref{table:notations}.

\begin{figure*}[ht]
\centering
\includegraphics[width=1\linewidth]{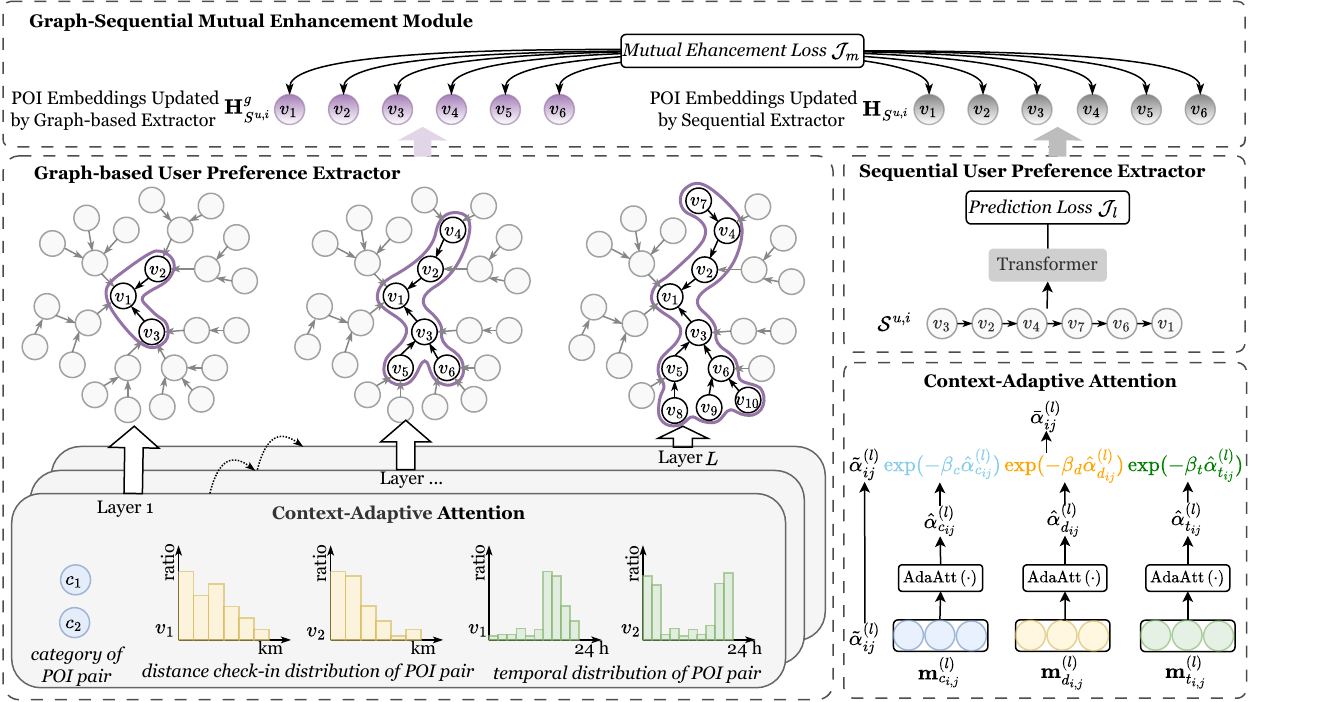}
\caption{The overall architecture of CAGNN.}
\label{fig:model_framework}
\vspace{-0.15in}
\end{figure*}

\smallskip\noindent\textbf{Model Overview.} Fig.~\ref{fig:model_framework} depicts the overall architecture of CAGNN, which is mainly composed of three modules. 
(1) \textit{Graph-based User Preference Extractor with Context-Adaptive Attention} jointly integrates multiple types of edge-specific contextual factors to dynamically adjust the attention weights in standard GAT during graph propagation. This context-adaptive mechanism enhances the modeling of collaborative and context-dependent patterns, thereby improving prediction accuracy.
(2) \textit{Sequential User Preference Extractor} focuses on capturing each user's sequential preferences.
(3) \textit{Graph-Sequential Mutual Enhancement Module} aligns the learned POI embeddings from graph-based and sequential user preference extractors, facilitating mutual enhancement to improve recommendation accuracy.

\subsection{Graph-based User Preference Extractor with Context-Adaptive Attention}

\smallskip\noindent\textbf{Representation Learning via GAT.}
Given the constructed directed POI graph, we adopt the commonly used GAT to learn POI representations by capturing collaborative information from the transition patterns of all users. Take one POI $v_i$ in $\mathcal{G}$ as an example, the feature propagation at the ($l$+1)-th layer is formulated as follows:
\begin{equation}
\small
\label{equ:GAT_prop}
\mathbf{h}_i^{(l+1)}=\sigma\left(\sum\nolimits_{j \in \mathcal{N}(i)} \alpha_{i j}^{(l)} \mathbf{W}^{(l)} \mathbf{h}_j^{(l)}\right),
\end{equation}
where $\mathbf{h}_i^{(l+1)}\in \mathbb{R}^{dim}$ is the representation of POI node $v_i$ at layer ($l+1$), $dim$ is the embedding size; 
$\mathbf{h}_j^{(l)} \in \mathbb{R}^{dim}$ is the representation of node $v_j$ at layer $l$, where $\mathbf{h}_j^{(0)}=\mathbf{v}_j$, and $\mathbf{v}_j \in \mathbf{V}$ is the initial POI embedding;
$\mathcal{N}(i)$ is the set of neighbors of POI node $v_i$; 
$\mathbf{W}^{(\boldsymbol{l})}\in \mathbb{R}^{dim * dim}$ is the learnable weight matrix at layer $l$;
$\sigma$ is an activation function (i.e., ReLU);
and $\alpha_{ij}^{(l)}$ is the attention weight that represents the importance of POI node $v_j$ to $v_i$ based on their embeddings at layer $l$, which is computed as:
\begin{equation}
\small
\label{equ:GAT_attention}
\begin{aligned}
    \tilde{\alpha}_{ij}^{(l)} &= \mathbf{a}^{(l)^T}\left[\mathbf{W}^{(l)} \mathbf{h}^{(l)}_i \| \mathbf{W}^{(l)} \mathbf{h}^{(l)}_j\right] \\
    \alpha^{(l)}_{i j}&=\frac{\exp \left(\operatorname{LeakyReLU}\left(\tilde{\alpha}_{ij}^{(l)}\right)\right)}{\sum_{k \in \mathcal{N}(i)} \exp \left(\operatorname{LeakyReLU}\left(\tilde{\alpha}^{(l)}_{i k}\right)\right)},
    \end{aligned}
\end{equation}
where $\tilde{\alpha}_{ij}^{(l)}$ is the unnormalized attention weight between POI nodes $v_i$ and $v_j$ based on their representations; 
$\mathbf{a}$ is a learnable attention vector;
$\|$ denotes vector concatenation;
and LeakyReLU is a nonlinear activation function.

\smallskip\noindent\textbf{Context-Adaptive Attention.}
As illustrated in Eqs.~(\ref{equ:GAT_prop}) and~(\ref{equ:GAT_attention}), GAT aggregates neighboring node features through attention-weighted summation based solely on node embeddings. However, user transitions between POIs (edges in the transition graph $\mathcal{G}$) are often influenced by multiple types of contextual factors~\cite{sun2021point}, such as POI category, time (e.g., peak hours), and spatial distance.
Ignoring these factors limits GAT’s ability to capture context-dependent transition patterns, resulting in suboptimal attention weight assignments.

To address this limitation, we propose a context-adaptive attention mechanism that dynamically adjusts standard GAT attention weights by jointly modeling the co-influence of multiple types of contextual factors during message propagation. Specifically, it comprises three learnable attention components that separately encode POI categories, spatial and temporal patterns. 
By incorporating edge-specific contextual factors, the mechanism assigns more accurate attention weights, highlighting relevant transitions while filtering out less meaningful ones. This enhances the modeling of context-dependent user behavior, resulting in more expressive POI representations and improved recommendation accuracy.

\textit{\textbf{(1) Category-aware Adaptive Attention.}} The POI category is one type of commonly used contextual factors in existing next POI recommendation~\cite{wang2024bi, li2024mcn4rec}, which denotes the user's intention. 
We concatenate the category features of POIs $v_i$ and $v_j$ to capture the categorical difference within the POI transition pair, 
\begin{equation} 
\small
\label{equ:mlp}
\begin{aligned}
\mathbf{m}^{(0)}_{c_{i,j}}&=[\mathbf{c}_i||\mathbf{c}_j]\\
\mathbf{m}^{(l)}_{c_{i,j}}&=\operatorname{ReLU}(\mathbf{W}^{(l-1)}_{c}\mathbf{m}^{(l-1)}_{c_{i,j}}+\mathbf{b}^{(l-1)}_c),
\end{aligned}
\end{equation}
\begin{equation}
\small
\label{equ:mlp_atten}
    \hat{\alpha}^{(l)}_{c_{i j}}=\operatorname{Mean}(\mathbf{m}^{(l)}_{c_{i,j}}),
\end{equation}
where $\mathbf{m}^{(0)}_{c_{i,j}} \in \mathbb{R}^{2*dim}$ is the input of the category-aware adaptive attention mechanism at the 0-th layer; 
$\mathbf{c}_i$ and $\mathbf{c}_j \in \mathbf{C}$ are the category embeddings of POI $v_i$ and $v_j$; 
$\mathbf{m}^{(l-1)}_{c_{i,j}}$ is the hidden output of the ($l$-$1$)-th layer and the input of the $l$-th layer;
$\mathbf{W}^{(l)}_{c} \in \mathbb{R}^{dim * dim}$ is a learnable weight matrix;
$\mathbf{b}^{(l)}_c \in \mathbb{R}^{dim}$ is a learnable bias; 
and $\hat{\alpha}^{(l)}_{c_{i j}}$ is the category-aware attention weight between $v_i$ and $v_j$ at the $l$-th layer. 
A larger value of $\hat{\alpha}^{(l)}_{c_{ij}}$ indicates greater disparity between their categories and is used to adjust the attention weights learned by standard GAT $\alpha^{(l)}_{ij}$ at the same layer.
For ease of illustration, we simplify Eqs. (\ref{equ:mlp}) and (\ref{equ:mlp_atten}) as below:
\begin{equation}
\small       
\label{equ:c_att}
    \hat{\alpha}^{(l)}_{c_{i j}}=\text{AdaAtt}([\mathbf{c}_i||\mathbf{c}_j]).
\end{equation}

\textit{\textbf{(2) Spatial-aware Adaptive Attention.} }
This module aims to capture the distance transition pattern of each edge, as users' behavior often exhibits strong spatial patterns~\cite{sun2023multi}, such as a preference for traveling shorter distances.
Existing approaches~\cite{zhao2020where,lim2020stp} often directly encode spatial information using the exact distance interval (commonly computed via the Haversine formula) between each POI node pair. However, these methods can be computationally expensive and challenging to learn due to the large variance in distance values. To better learn the spatial difference between each POI node pair, we utilize the spatial check-in distribution of each POI to represent its spatial feature. Specifically, we differentiate a POI as either a neighbor or a central node in a POI node pair, as users' movement patterns may vary depending on whether they are departing from or arriving at a location. This distinction enables us to capture directional spatial dependencies more effectively. The formulation is as follows:
\begin{equation} 
\small
\begin{aligned}
\mathbf{d}^{src}_{i}&=\mathbf{d}_{i,:}*\mathbf{D},\\
\mathbf{d}^{dst}_{i}&=\mathbf{d}_{:,i}*\mathbf{D},
\end{aligned}
\end{equation}
where $\mathbf{D}$$\in$$\mathbb{R}^{\operatorname{max}(\mathcal{D})*dim}$ is the distance embedding matrix; $\mathbf{d}_{i,:}$$\in$$\mathbb{R}^{\operatorname{max}(\mathcal{D})}$ represents the check-in distribution for each distance interval when POI $v_i$ is the neighbor node, with each element corresponding to the transition ratio for a specific distance interval; 
and $\mathbf{d}_{:,i}$ records the check-in distribution for each distance interval when POI $v_i$ is the central node.

After obtaining the spatial features of each POI, similar to the category-aware adaptive attention mechanism, we concatenate the spatial features of each POI node pair as input to compute the spatial-aware attention weights at each layer, capturing the spatial disparity between POIs. A high value of $\hat{\alpha}^{(l)}_{d_{ij}}$ indicates that the distance between POIs $v_i$ and $v_j$ is uncommon for user transitions, suggesting that the edge is less likely to represent a meaningful relationship,
\begin{equation}
\small       
\label{equ:d_att}
    \hat{\alpha}^{(l)}_{d_{i j}}=\text{AdaAtt}([\mathbf{d}^{src}_{i}||\mathbf{d}^{dst}_{j}]).
\end{equation}
%
%

\textit{\textbf{(3) Temporal-aware Adaptive Attention.}}
As our goal is to predict the POIs that users are likely to visit at a given time, it is vital to learn the temporal features of each POI node pair.
In general, POIs are visited during specific time slots. For example, POIs categorized as `food' are often visited during typical meal times, such as 11 AM–1 PM and 5 PM–7 PM.
To capture this temporal pattern, we represent each POI's temporal feature using its hourly check-in frequency over the 24 hours of the day:
\begin{equation} 
\small
    \hat{\mathbf{t}}^{v_i}=\mathbf{t}^{v_i}*\mathbf{T},
\end{equation}
where $\mathbf{T} \in \mathbb{R}^{24*dim}$ is the timeslot embedding matrix; 
and $\mathbf{t}^{v_i} \in \mathbb{R}^{24}$ denotes the hourly check-in frequency of POI $v_i$ in each hour of the day. Then, we use the temporal representations of each POI node pair as input to learn their temporal disparity,
\begin{equation}
\small      
\label{equ:t_att}
    \hat{\alpha}^{(l)}_{t_{i j}}=\text{AdaAtt}([\hat{\mathbf{t}}^{v_i}||\hat{\mathbf{t}}^{v_j}]).
\end{equation}
%
%
%
\smallskip\noindent\textbf{Propagation with Context-Adaptive Attention.} 
As shown in the GAT work~\cite{velivckovic2018graph}, selectively aggregating information from a central node’s neighbors is effective for node representations learning. 
However, standard GAT computes edge attention weights solely based on node embeddings and thus overlooks edge-specific contextual factors that may affect the relative importance of each neighbor during propagation.
To address this, we integrate the proposed context-adaptive attention mechanism into the GAT’s forward propagation.
Since the raw context-aware attention weights $\hat{\alpha}^{(l)}_{c_{ij}}$, $\hat{\alpha}^{(l)}_{d_{ij}}$, and $\hat{\alpha}^{(l)}_{t_{ij}}$ can be either positive or negative, reflecting varying degrees of contextual disparity between nodes, we apply the exponential function to their negatives, as shown in Equ.~(\ref{equ:att_context}). Leveraging the non-negativity of the exponential function, this operation transforms the raw values into non-negative and bounded attention weights. Consequently, neighbors with higher raw weights (indicating greater contextual disparity) yield smaller exponential values, reducing their influence during message passing. In contrast, neighbors with lower raw weights produce larger exponential values, thereby increasing their contribution to the central node’s representation.
This mechanism enables the model to focus more on influential neighbors and suppress less meaningful ones, improving the quality of learned node representations.

The details of the context-adaptive attention are formulated as follows:
\begin{equation}
\label{equ:att_context}
    \small
    \hat{\alpha}^{(l)}_{ij} = exp(-\beta_c\hat{\alpha}^{(l)}_{c_{i j}})exp(-\beta_d\hat{\alpha}^{(l)}_{d_{i j}})exp(-\beta_t\hat{\alpha}^{(l)}_{t_{i j}}),
\end{equation}
where $\hat{\alpha}^{(l)}_{ij}$ the context-adaptive attention weight between $v_i$ and $v_j$ at layer $l$ during propagation; 
$\beta_c$, $\beta_d$, and $\beta_t$ are learnable parameters to balance the influence of the three types of context, with $\beta_c$+$\beta_d$+$\beta_t$=$1$;
and $\hat{\alpha}^{(l)}_{c_{i j}}$, $\hat{\alpha}^{(l)}_{d_{i j}}$ and $\hat{\alpha}^{(l)}_{t_{i j}}$ are adaptive attention weights learned by Eqs.~(\ref{equ:c_att}),~(\ref{equ:d_att}), and~(\ref{equ:t_att}) at layer $l$.
The normalized attention weights between POI nodes $v_i$ and $v_j$ in Equ.~(\ref{equ:GAT_attention}) and the feature propagation of $v_i$ at the ($l$+1)-th layer in Equ.~(\ref{equ:GAT_prop}) can be reformulated as follows:
\begin{equation}\label{equ:GAT_m}
    \small
    \begin{aligned}
    &\bar{\alpha}^{(l)}_{i j}=\frac{\exp \left(\operatorname{LeakyReLU}\left(\tilde{\alpha}_{ij}^{(l)}\hat{\alpha}^{(l)}_{ij}\right)\right)}{\sum_{k \in \mathcal{N}(i)} \exp \left(\operatorname{LeakyReLU}\left(\tilde{\alpha}_{ik}^{(l)}\hat{\alpha}^{(l)}_{ik}\right)\right)},\\
    &\mathbf{h}_i^{(l+1)}=\sigma\left(\sum\nolimits_{j \in \mathcal{N}(i)} \bar{\alpha}^{(l)}_{i j} \mathbf{W}^{(l)} \mathbf{h}_j^{(l)}\right).
\end{aligned}
\end{equation}
Accordingly, the representation of POI $v_i$, updated through GAT propagation with context-adaptive attention mechanism on the POI–POI transition graph, is given by:
\begin{equation}
\label{equ:graph_rep}
\small
     \mathbf{h}_i^{g} = \frac{1}{|L|}\left(\sum\nolimits_{l \in L} \mathbf{h}_i^{(l)}\right),
\end{equation}
where $\mathbf{h}_i^{g} \in \mathbf{H}^{g}$ is the updated embedding of POI $v_i$, and $\mathbf{H}^{g} \in \mathbb{R}^{|\mathcal{V}|\times dim}$ represents the POI embedding matrix after graph propagation; 
and $L$ denotes the number of layers in the GAT. The overall process of graph-based user preference extractor with context-adaptive attention mechanism is described in Algorithm~\ref{algo:1}.

\setlength{\textfloatsep}{0pt}
\begin{algorithm}[t]
    \footnotesize
    \caption{\small Pseudocode for Graph-based User Preference Extractor with Context-Adaptive Attention Mechanism}
    \label{algo:1}
    \KwIn {$\mathcal{G}$, $\mathbf{V}$, $\mathbf{C}$, $\mathbf{D}$, $\mathbf{T}$;}
    \For{($l=1; iter \leq L; l++$)}
    {
        $\tilde{\alpha}_{ij}^{(l)}\leftarrow$ Calculate node embedding-based attention of each POI pair ($v_i, v_j$) according to Equ.~(\ref{equ:GAT_attention})\;
        $\hat{\alpha}^{(l)}_{c_{i j}}\leftarrow$ Calculate category-aware adaptive attention of each POI pair ($v_i, v_j$) according to Equ.~(\ref{equ:c_att})\;
        $\hat{\alpha}^{(l)}_{d_{i j}}\leftarrow$ Calculate spatial-aware adaptive attention of each POI pair ($v_i, v_j$) according to Equ.~(\ref{equ:d_att})\;
        $\hat{\alpha}^{(l)}_{t_{i j}}\leftarrow$ Calculate temporal-aware adaptive attention of each POI pair ($v_i, v_j$) according to Equ.~(\ref{equ:t_att})\;
        $\hat{\alpha}^{(l)}_{i,j}\leftarrow$ Calculate context-sensitive adaptive attention of each POI pair ($v_i, v_j$) according to Equ.~(\ref{equ:att_context})\;
        Graph propagation according to Equ.~(\ref{equ:GAT_m})
    }
    \KwOut{$\mathbf{H}^{g}$}
\end{algorithm}

\subsection{Sequential User Preference Extractor}
This module aims to capture sequential dependencies based on the check-in sequence $\mathcal{S}^{u, i}$. 
We employ a bidirectional Transformer~\cite{vaswani2017attention} as the encoder due to its strong representation learning capability for sequential data.
Given a user $u$'s check-in sequence $\mathcal{S}^{u, i}=\{S_{t_1}^u, S_{t_2}^u,\cdots, S_{t_k}^u\}$,  $S_{t_k}^u = (u, v_{t_k}^u, t_{k}^u)$ can be represented as:
\begin{equation}
\small
\label{equ:cat}
    \mathbf{e}_{\tilde{v}_{t_k}^u} = [\mathbf{u}\|\mathbf{v}_{t_k}^u\|\mathbf{t}_{k}],
\end{equation}
where $\mathbf{e}_{\tilde{v}_{t_k}^u} \in \mathbb{R}^{3\times dim}$;  $\mathbf{u}, \mathbf{v}_{t_k}^u, \mathbf{t}_{k}\in \mathbb{R}^{dim}$ are the embeddings of user $u$, POI $v_{t_k}^u$, and time slot $t_{k}^u$, respectively. 
Accordingly, the representation of $\mathcal{S}^{u, i}$ is $\mathbf{E}_{S^{u, i}} = [\mathbf{e}_{\tilde{v}_{t_1}^u}, \mathbf{e}_{\tilde{v}_{t_2}^u},\cdots, \mathbf{e}_{\tilde{v}_{t_k}^u}]$. Then, we feed it to a transformer layer:
\begin{equation}
\small
\label{equ:trans_output}
\begin{aligned}
    \mathbf{H}_{S^{u, i}} &= [\mathbf{h}_{\tilde{v}_{t_1}^u}, \mathbf{h}_{\tilde{v}_{t_2}^u},\cdots, \mathbf{h}_{\tilde{v}_{t_k}^u}] \\
    &=\text{Trans}(\mathbf{E}_{S^{u, i}}\mathbf{W}_Q, \mathbf{E}_{S^{u, i}}\mathbf{W}_K, \mathbf{E}_{S^{u, i}}\mathbf{W}_V) ,
    \end{aligned}
\end{equation}
\begin{equation}
\small
\label{equ:trans}
    \text{Trans}(\mathbf{Q},\mathbf{K},\mathbf{V}) = (softmax(\frac{\mathbf{Q}\mathbf{K}^\intercal}{\sqrt{3\times dim}}))\mathbf{V},
\end{equation}
where $\mathbf{H}_{S^{u, i}}$ is the output matrix of the transformer layer, representing the updated embedding matrix of POIs in sequence $S^{u, i}$ after the transformer layer; 
$\text{Trans}(\cdot)$ denotes a transformer layer; 
$\mathbf{h}_{\tilde{v}_{t_1}^u}, \mathbf{h}_{\tilde{v}_{t_2}^u},\cdots, \mathbf{h}_{\tilde{v}_{t_k}^u}\in \mathbb{R}^{3\times dim}$ represent the updated POI embeddings; 
and $\mathbf{W}_Q$, $\mathbf{W}_K$, and $\mathbf{W}_V\in \mathbb{R}^{3dim\times 3dim}$ are the corresponding learnable weight matrices of query, key, and value for the transformer layer.

\subsection{Graph-Sequential Mutual Enhancement Module}
This module aims to align the learned POI embeddings from both the graph-based and sequential user preference extractors to mutually enhance their learning process. The sequential component excels at capturing sequential patterns in user behavior but cannot incorporate collaborative information from other users, which is particularly beneficial for inactive users. 
On the contrary, the graph-based component effectively learns transition patterns across users through information propagation, but struggles to capture sequential dependencies within a user’s trajectory.
Existing approaches~\cite{rao2025next,li2024mcn4rec,wang2023adaptive} often treat sequential models (e.g., LSTM or Transformer) as the backbone of the prediction model and feed POI embeddings learned via the graph-based component into them as prior knowledge. 
As such, the sequential component may dominate the prediction process, potentially undermining the semantic and structural information encoded in the POI embeddings learned by GNNs.

Our approach addresses this by aligning POI embeddings from both components. 
Specifically, we first obtain the embedding of POIs in the sequence $S^{u, i}$ learned by the two types of extractors (Eqs.(\ref{equ:graph_rep}) and (\ref{equ:trans_output})) and compute the distribution of each POI embedding:
\begin{equation}
\small
\begin{aligned}
     \operatorname{Softmax}(\mathbf{H}^g_{S^{u, i}}) &= \operatorname{Softmax}([\mathbf{h}^{g}_{v_{t_1}^u}, \mathbf{h}^{g}_{v_{t_2}^u}, \cdots, \mathbf{h}^g_{v_{t_k}^u}]),\\
     \operatorname{Softmax}(\mathbf{H}_{S^{u, i}}) &= \operatorname{Softmax}([\mathbf{h}_{\tilde{v}_{t_1}^u}, \mathbf{h}_{\tilde{v}_{t_2}^u},\cdots, \mathbf{h}_{\tilde{v}_{t_k}^u}]),\\
\end{aligned}
\end{equation}
where $\operatorname{Softmax}(\mathbf{H}^g_{S^{u, i}})$ denotes the distribution of POI embeddings in the sequence $S^{u, i}$ after graph propagation in the graph-based extractor; 
$\operatorname{Softmax}(\mathbf{H}_{S^{u, i}})$ denotes the distribution of POI embeddings in the sequence $S^{u, i}$ learned by the sequential extractor; 
and $\operatorname{Softmax(\cdot)}$ is the softmax function.
Then, we align the two distributions to mutually influence both components during back propagation:
\begin{equation}
\small
\begin{aligned}
    \mathcal{J}^{sg}_{m} = \frac{1}{|\mathcal{S}|}\sum_{S^{u, i} \in\mathcal{S}}\operatorname{KL}&[\operatorname{Softmax}(\mathbf{H}_{S^{u, i}}) \|\operatorname{Softmax}(\mathbf{H}^g_{S^{u, i}})],\\
    \mathcal{J}^{gs}_{m} = \frac{1}{|\mathcal{S}|}\sum_{S^{u, i} \in\mathcal{S}}\operatorname{KL}&[\operatorname{Softmax}(\mathbf{H}^g_{S^{u, i}}) \|\operatorname{Softmax}(\mathbf{H}_{S^{u, i}})],
\end{aligned}
\end{equation}
\begin{equation}
    \small
    \operatorname{KL}[P_1\|P_2] = \sum\nolimits^{t_k}_{t=t_1}P_1(t)\operatorname{log}(\frac{P_1(t)}{P_2(t)}),
\end{equation}
\begin{equation}
    \small
    \mathcal{J}_{m} = \mathcal{J}^{sg}_{m} + \mathcal{J}^{gs}_{m},
\end{equation}
where $\mathcal{J}^{sg}_{m}$ and $\mathcal{J}^{gs}_{m}$ bi-directionally align the two distributions; 
$|\mathcal{S}|$ represents the number of training sequences; 
and $\operatorname{KL}[\cdot]$ represents the Kullback-Leibler (KL) divergence~\cite{kullback1951information}.

\subsection{Model Training and Complexity Analysis}
\smallskip\noindent{\textbf{POI Prediction.}} To improve personalization, we represent the user's current preference with the embedding of their most recently visited POI, generated by the sequential extractor, i.e., $\mathbf{h}_{\tilde{v}_{t_k}^u}$. The learned sequential user preference $\mathbf{h}_{\tilde{v}_{t_k}^u}$ is then used to decode the probability distribution over the $|\mathcal{V}|$ POIs using the softmax function:
\begin{equation}
    \small
    \hat{\mathbf{y}}_{t_{k+1}} = \operatorname{Softmax}(\mathbf{W}_s\mathbf{h}_{\tilde{v}_{t_k}^u} + \mathbf{b}_s),
\end{equation}
where $\mathbf{W}_s \in \mathbb{R}^{3dim\times|\mathcal{V}|}$, $\mathbf{b}_s \in \mathbb{R}^{|\mathcal{V}|}$.
Hence, the cross-entropy loss function for the next POI prediction is defined as:
\begin{equation}
\small
\mathcal{J}_l= -\frac{1}{|\mathcal{S}|}\sum_{\mathcal{S}} \text{log} (\hat{\mathbf{y}}_{t_{k+1}}).
\end{equation}
\smallskip\noindent{\textbf{Final Objective Function.}} Ultimately, the overall loss function of CAGNN is as follows:
\begin{equation}
\small
\mathcal{J}= \mathcal{J}_l + \beta\mathcal{J}_{m} + \lambda||\theta||^2_2,
\end{equation}
where $\beta$ is a hyperparameter to balance the two objective functions $\mathcal{J}_l$ and $\mathcal{J}_{m}$; $\theta$ represents all trainable model parameters, and $\lambda$ is the coefficient for the L2 regularization term. 

\smallskip\noindent{\textbf{Complexity Analysis.}} We discuss the time complexity of the three modules in CAGNN separately. For the graph-based user preference extractor, the complexity consists of two parts. First, computing the context-adaptive attention requires training three attention modules across $L$ layers. The input of each module is the concatenation of the feature representations for POIs in each edge, which has a complexity of $\mathcal{O}(L\cdot|\mathcal{E}|\cdot dim)$. 
Second, updating node representations via GAT over the POI-POI transition graph $\mathcal{G} = (\mathcal{V}, \mathcal{E})$ also has a complexity of $\mathcal{O}(L \cdot |\mathcal{E}| \cdot dim)$, as attention weights are computed for each edge at every layer.
For the sequential user preference extractor, we need to train a transformer for each training sequence with a maximum length of 
$\text{max}(|\mathcal{S}^{u,i}|)$. The computational complexity is $\mathcal{O}(\text{max}(|\mathcal{S}^{u,i}|)^2\cdot dim)$.
Last, for the graph-sequential mutual enhancement module, the KL divergence is computed for each training sequence; the total complexity is $\mathcal{O}(\text{max}(|\mathcal{S}^{u,i}|)\cdot dim)$.
In practice, the number of transition edges in the POI–POI graph is larger than the number of POI nodes, satisfying $L < \text{max}(|\mathcal{S}^{u,i}|) < dim < \vert\mathcal{V}\vert < \vert\mathcal{E}\vert$. Thus, the overall complexity of the model for each training sequence is approximately $\mathcal{O}(L \cdot \vert\mathcal{E}\vert \cdot \text{dim})$, where $\vert\mathcal{E}\vert$ may vary depending on the dataset, which implies that the model is scalable to large-scale datasets as long as the graph size is manageable.
\section{Expressiveness Analysis}\label{sec:proof}
The context-adaptive attention mechanism plays a key role in CAGNN. It jointly incorporates multiple types of contextual information to model their co-influence on user transitions, thereby enhancing graph propagation and facilitating the learning of more expressive POI representations.
To demonstrate the superior capability of the context-adaptive attention mechanism, we conduct a qualitative analysis comparing it to standard GAT. Specifically, we analyze the differences based on Equ.~(\ref{equ:att_context}) and~(\ref{equ:GAT_m}) for GAT with context-adaptive attention mechanism and Eqs.~(\ref{equ:GAT_prop}) and~(\ref{equ:GAT_attention}) for standard GAT.

\begin{theorem}
Let $\psi (\cdot)$ be a learnable function for node representation learning used in the context-adaptive attention mechanism and standard GAT methods. Let $V = \{V_i|i = 1, ..., n\}$ and $C = \{C_i|i = 1, ..., n\}$ denote the sets of correlation information derived from POI embeddings and contextual factors (i.e., category, spatial and temporal information), respectively, which are used for learning node representations in a graph. Additionally, let $Y = \{Y_i|i = 1, ..., n\}$ represent the set of ground-truth labels of predictions, where $n$ denotes the number of samples used for training. The node representations learned by the GAT with context-adaptive attention mechanism and standard GAT are denoted as $\bar{z}_i = \psi (V_i \cdot C_i)$ and $z_i = \psi (V_i)$, respectively, for $i = 1, ..., n$. Then, for any $i \in {1, ..., n}$, the following inequality holds:
\begin{equation}
\small
I (Y_i; \bar{z}_i ) \geq I(Y_i; z_i ), 
\end{equation}
where $I(\cdot; \cdot)$ denotes the mutual information function\footnote{The POI prediction task in our model depends not only the POI embeddings updated in the graph, but for ease of illustration, we focus solely on the relationship between the prediction labels $Y_i$ and the POI node representations learned by the graph ($\bar{z}_i$ and ${z}_i$)}.
\end{theorem}

\begin{customproof}
To prove Theorem 1, we utilize the following properties of mutual information and entropy:
\begin{equation}
\small
    \begin{aligned}
        &I(x; y) = H(y) - H(y|x),  \\
        &I(x; z|y) = H(x|y) - H(x|y, z), \\ 
        &H(x, y) = H(x|y) + H(y),
    \end{aligned}
\end{equation}
where $I(\cdot; \cdot)$ denotes the mutual information between random variables $x$, $y$, and $z$, $H(\cdot)$ represents marginal entropy, and $H(\cdot|\cdot)$ represents conditional entropy.
Let $I(Y_i; \bar{z}_i)$ denote the mutual information between the ground-truth label $Y_i$ and the representation $\bar{z}_i = \psi (V_i \cdot C_i)$ learned by the GAT with the context-adaptive attention mechanism. Based on Equation~(\ref{equ:GAT_m}), the model first captures the joint effect of $V_i$ and $C_i$, then applies the shared function $\psi(\cdot)$ to generate node representations. This process is equivalent to directly leveraging both $V_i$ and $C_i$ to formulate dual messages for representation learning. Consequently, $I(Y_i; \psi (V_i \cdot C_i))$ can be rewritten as $I(Y_i; \psi (V_i), \psi (C_i))$, where $\psi (V_i)$ and $\psi (C_i)$ are drawn from a joint distribution. Thus, we derive:
\begin{equation}
\small
    \begin{aligned}
        &I(Y_i; \psi (V_i), \psi (C_i)) = H(\psi (V_i), \psi (C_i)) - H(\psi (V_i), \psi (C_i)|Y_i),  \\
        &= H(\psi (V_i), \psi (C_i)) + H(Y_i) - H(\psi (V_i), \psi (C_i), Y_i) \\ 
        &= H(Y_i) - H(Y_i|\psi (V_i), \psi (C_i))  \\ 
        &= H(Y_i) - H(Y_i|\psi (V_i)) + H(Y_i|\psi (V_i)) - H(Y_i|\psi (V_i), f (C_i))  \\
        &= I(Y_i; \psi (V_i)) + I(Y_i; \psi (C_i)|\psi (V_i)) \geq I(Y_i; \psi (V_i)), 
    \end{aligned}
\end{equation}
where $I(Y_i; \psi (V_i))$ represents the mutual information between $Y_i$ and the representation learned by standard GAT $z_i = \psi (V_i)$. As mutual information is non-negative, the above inequality holds for any function $\psi (\cdot)$. Moreover, since graph data often exhibit structural dissimilarities and both $V_i$ and $C_i$ are derived from the same set of nodes, $I(Y_i; \psi (C_i)|\psi (V_i))$ is likely greater than zero. This results in a larger gap between $I(Y_i; \psi (C_i), \psi (V_i))$ and $I(Y_i; \psi (V_i))$ for any node in the graph, reinforcing the advantage of incorporating contextual information in attention modulation.
\end{customproof}

The proof of Theorem 1 demonstrates that the context-adaptive attention mechanism enables GAT to consistently learn more expressive POI node embeddings compared to the standard GAT. These enhanced embeddings contribute to improved POI recommendation accuracy when compared to standard GAT, as long as $\psi (\cdot)$ is well defined.
\section{Experiment}
\begin{table}[t]
\centering
\caption{Statistics of the three datasets used in our study.}\label{tab:data_statistic}
\vspace{-0.1in}
  \addtolength{\tabcolsep}{4pt}
  \begin{tabular}{l|rrrc}
    \specialrule{.15em}{.15em}{.15em}
    Dataset&\#Users&\#POIs &\#Check-ins &Density\\
    \specialrule{.1em}{.1em}{.1em}
    PHO&2,946&7,247 &47,980 & 0.22\%\\
    NY&16,387&56,252 &511,431 & 0.06\%\\
    SIN&8,648 &33,712 &355,337 & 0.12\%\\
  \specialrule{.15em}{.15em}{.15em}
\end{tabular}
\end{table}
In this section, we present the experimental setup and conduct comprehensive empirical evaluations on three real-world datasets to assess the effectiveness of the proposed CAGNN. We begin by comparing the performance of CAGNN with several SOTA models. Next, we investigate the contributions of different components within CAGNN and analyze the impact of key hyperparameters. Finally, we examine the distribution of attention scores produced by the context-adaptive attention mechanism in comparison to standard GAT, and provide a case study to illustrate the attention generated by the context-adaptive attention mechanism.

\begin{table*}[ht]
\centering
\caption{Performance of all methods on the three datasets. The performance of CAGNN is boldfaced; the performance of the best baseline is underlined; and ‘Improve' shows improvements achieved by CAGNN relative to the best baseline, whose significance is determined by a paired t-test (** for $p<0.01$ and *** for $p<0.001$).}
\label{tab:comparision}
\addtolength{\tabcolsep}{-1.5pt}
\begin{tabular}{c|l|c|c|c|c|c|c|c|c|c|c|c|c}
\specialrule{.15em}{.05em}{.05em}
\multicolumn{2}{c|}{} & \multicolumn{3}{c|}{RNN-based} & \multicolumn{3}{c|}{Transformer-based} & \multicolumn{4}{c|}{Graph-based} &  \multicolumn{2}{c}{} \\
\specialrule{.05em}{.05em}{.05em}
\multicolumn{2}{c|}{Metric} & ST-RNN & ATST-LSTM & PLSPL & CTLE & CFPRec & CLSPRec & LightGCN & SGRec & AGRAN & DCHL & CAGNN & Improve \\
\specialrule{.1em}{.05em}{.05em}
\multirow{6}{*}{PHO} & HR@1 & 0.0419 & 0.0517 & 0.0846 & 0.1439 & 0.1842 & \underline{0.2521} & 0.0775 & 0.1295 & 0.1937 & 0.2057 & \textbf{0.2945}*** & 16.82\% \\
& HR@5 & 0.1240 & 0.1579 & 0.1775 & 0.2632 & 0.3421 & \underline{0.5217} & 0.2563 & 0.2897 & 0.4573 & 0.4962 & \textbf{0.5611}** & 7.55\% \\
& HR@10 & 0.2028 & 0.2377 & 0.2569 & 0.3605 & 0.4253 & \underline{0.5936} & 0.3151 & 0.3401 & 0.5248 & 0.5714 & \textbf{0.6271}** & 5.64\% \\
& N@1 & 0.0419 & 0.0517 & 0.0846 & 0.1439 & 0.1842 & \underline{0.2521} & 0.0775 & 0.1295 & 0.1937 & 0.2057 & \textbf{0.2945}*** & 16.82\% \\
& N@5 & 0.0802 & 0.1033 & 0.1285 & 0.1995 & 0.2432 & \underline{0.3806} & 0.1881 & 0.2048 & 0.3685 & 0.3336 & \textbf{0.4347}*** & 14.21\% \\
& N@10 & 0.1229 & 0.1385 & 0.1538 & 0.2068 & 0.2730 & \underline{0.4126} & 0.2194 & 0.2249 & 0.3962 & 0.3638 & \textbf{0.4566}*** & 10.66\% \\
\specialrule{.05em}{.05em}{.05em}
\multirow{6}{*}{NY} & HR@1 & 0.0327 & 0.0396 & 0.0426 & 0.0649 & 0.0718 & \underline{0.1797} & 0.0527 & 0.0674 & 0.1296 & 0.1498 & \textbf{0.2065}*** & 14.91\% \\
& HR@5 & 0.1347 & 0.1667 & 0.1741 & 0.2421 & 0.2771 & \underline{0.3518} & 0.1752 & 0.1891 & 0.2732 & 0.3292 & \textbf{0.4023}*** & 13.45\% \\
& HR@10 & 0.1826 & 0.2031 & 0.2413 & 0.3205 & 0.3606 & \underline{0.4388} & 0.2229 & 0.2443 & 0.3746 & 0.3904 & \textbf{0.4880}*** & 11.21\% \\
& N@1 & 0.0327 & 0.0396 & 0.0426 & 0.0649 & 0.0718 & \underline{0.1797} & 0.0527 & 0.0674 & 0.1296 & 0.1498 & \textbf{0.2065}*** & 14.91\% \\
& N@5 & 0.0593 & 0.0912 & 0.0961 & 0.1513 & 0.1971 & \underline{0.2583} & 0.1035 & 0.1089 & 0.2097 & 0.2217 & \textbf{0.3076}*** & 19.09\% \\
& N@10 & 0.1303 & 0.1638 & 0.1825 & 0.1841 & 0.2190 & \underline{0.2792}& 0.1209 & 0.1877 & 0.2308 & 0.2457 & \textbf{0.3423}*** & 22.60\% \\
\specialrule{.05em}{.05em}{.05em}
\multirow{6}{*}{SIN} & HR@1 & 0.0439 & 0.0478 & 0.0513 & 0.0722 & 0.0744 & \underline{0.2132} & 0.0823 & 0.0921 & 0.1552 & 0.1732 & \textbf{0.2312}** & 8.47\% \\
& HR@5 & 0.0959 & 0.1296 & 0.1447 & 0.2041 & 0.2310 & 0.3564 & 0.2165 & 0.2310 & 0.2955 & \underline{0.3573} & \textbf{0.3890}** & 8.87\% \\
& HR@10 & 0.1370 & 0.1933 & 0.1719 & 0.2784 & 0.3085 & \underline{0.4077} & 0.2691 & 0.2953 & 0.3339 & 0.4045 & \textbf{0.4375}** & 7.31\% \\
& N@1 & 0.0439 & 0.0478 & 0.0513 & 0.0722 & 0.0744 & \underline{0.2132} & 0.0823 & 0.0921 & 0.1552 & 0.1732 & \textbf{0.2312}** & 8.47\% \\
& N@5 & 0.0655 & 0.1027 & 0.1126 & 0.1315 & 0.1588 & \underline{0.2763} & 0.1263 & 0.1530 & 0.2576 & 0.2670 & \textbf{0.3155}** & 14.20\% \\
& N@10 & 0.0794 & 0.1476 & 0.1384 & 0.1556 & 0.1836 & \underline{0.2939} & 0.1335 & 0.1739 & 0.2697 & 0.2856 & \textbf{0.3313}** & 12.74\% \\
\specialrule{.15em}{.05em}{.05em}
\end{tabular}
\vspace{-0.1in}
\end{table*}

\subsection{Experimental Setup}
\smallskip\noindent\textbf{Datasets}.
Following~\cite{zhang2022next, duan2023clsprec}, we use Foursquare~\cite{yang2016participatory} check-in records in three cities, namely, Phoenix (PHO), New York (NY), and Singapore (SIN) from April 2012 to September 2013. Then, we filter out POIs with fewer than 10 interactions, inactive users with fewer than 5 trajectories, and trajectories with fewer than 3 check-in records~\cite{zhang2022next}. For each user, we take the earlier 80\% of his trajectories as the training set; the latest 10\% of trajectories as the test set, and the rest 10\% as the validation set.
The statistics of the three datasets are summarized in Table~\ref{tab:data_statistic}.

\smallskip\noindent\textbf{Compared Baselines.}
We compare our model with three types of baselines.
The first type is the RNN-based baselines for next POI recommendation, including,
\textbf{ST-RNN}~\cite{liu2016predicting} which incorporates spatial and temporal contextual factors into an RNN for next POI prediction.;
\textbf{ATST-LSTM}~\cite{huang2019attention} which integrates spatial and temporal contextual factors in a multi-modal manner and applies an attention mechanism for next POI prediction;
\textbf{PLSPL}~\cite{wu2020personalized} which utilizes LSTM and attention to capture users' current and historical preferences.

The second type is the Transformer-based POI recommendation methods, including, 
\textbf{CTLE}~\cite{lin2021pre} which uses a bidirectional Transformer to learn contextual neighbors for next POI recommendation;
\textbf{CFPRec}~\cite{zhang2022next} which derives multi-step future preferences from history to enhance user preference modeling;
\textbf{CLSPRec}~\cite{duan2023clsprec} which leverages contrastive learning and Transformers to enhance user preference learning.

The third type is the graph-based method, as our proposed model is built upon the graph neural networks, including, \textbf{LightGCN}~\cite{he2020lightgcn} which is a simplified version of GCN without feature transformation or nonlinear activation for general item recommendation;
\textbf{SGRec}~\cite{li2021discovering} which uses graph-augmented sequences and category-aware multi-task learning for POI prediction;
\textbf{AGRAN}~\cite{wang2023adaptive} which adaptively constructs POI graphs via geographical similarity and integrates spatiotemporal features using self-attention;
\textbf{DCHL}~\cite{lai2024disentangled} which employs multi-view disentangled hypergraph and contrastive learning to model diverse user preferences.

\smallskip\noindent\textbf{Evaluation Metrics.}
Following SOTAs~\cite{li2021discovering}, Hit Rate (HR) and Normalized Discounted Cumulative Gain (NDCG) are adopted to evaluate the model performance regarding accuracy. Generally, higher metric values indicate better ranking accuracy.

\smallskip\noindent\textbf{Hyper-parameter Settings.}
We perform hyperparameter optimization via grid search based on NDCG@10 on the validation set~\cite{sun2020we}. The embedding size is selected from \{60, 90, 120, 150, 180\}; learning rate from \{$5e^{-5}, 1e^{-4}, 5e^{-4}, 1e^{-3}, 5e^{-3}$\}; alignment loss weight from \{0.1, 0.3, 0.7, 1, 3, 7, 10\}; and the number of GAT and Transformer layers from \{1, 2, 3, 4, 5\}. Other parameters are tuned following the original papers.
We implement our model in PyTorch with the Adam optimizer. For PHO/NY/SIN, we set the embedding size to 180/150/150, learning rate to $1e^{-4} / 1e^{-4} / 5e^{-5}$, the weight of mutual augmentation loss $\mathcal{J}_m$ to 0.7/1/1; and use 5/2/2 Transformer layers and 3/5/2 GAT layers, respectively. Training runs for 100 epochs with early stopping.

\subsection{Comparative Results and Analysis}\label{subsec:results-and-analysis}
The performance of all compared methods in terms of HR@K and NDCG@K (K = {1, 5, 10}) is detailed in Table~\ref{tab:comparision}. Each method was executed 5 times, and the average results are reported to ensure robustness. Several key observations can be drawn from the results.

\textbf{(1)} Among the RNN-based baselines, PLSPL performs best, followed by ATST-LSTM and ST-RNN. This highlights the importance of capturing users' sequential patterns and the effectiveness of attention mechanisms in sequential modeling.
\textbf{(2)} Transformer-based models (CTLE, CFPRec, CLSPRec) consistently outperform RNN-based baselines, demonstrating the superior capability of transformers in modeling complex sequential patterns.
\textbf{(3)} Among GNN-based methods, LightGCN and SGRec build predefined graphs from user check-in trajectories. Although LightGCN is a non-sequential model, it outperforms several RNN-based baselines, demonstrating the effectiveness of modeling collaborative signals via graph structures. SGRec further integrates sequential modeling and POI category information, achieving better performance than LightGCN. This suggests that incorporating contextual factors, such as POI categories, improves recommendation accuracy.
\textbf{(4)} Advanced GNN models that use adaptive graph structures and multi-context hypergraphs (e.g., AGRAN and DCHL) outperform simpler GNN variants like LightGCN and SGRec. This highlights the benefits of graph refinement and contextual modeling. However, these models occasionally underperform compared to SOTA transformer-based methods, indicating the need for more effective integration of graph and sequential information.
\textbf{(5)} Our proposed CAGNN consistently achieves the best performance across all metrics. On average, it outperforms the runner-up by 10.47\% in HR and 14.85\% in NDCG across the three datasets, demonstrating its effectiveness.

\begin{figure*}[t]
\centering
\graphicspath{{figure/}} 
\begin{minipage}{0.3\linewidth}
  \centerline{\includegraphics[width=5cm]{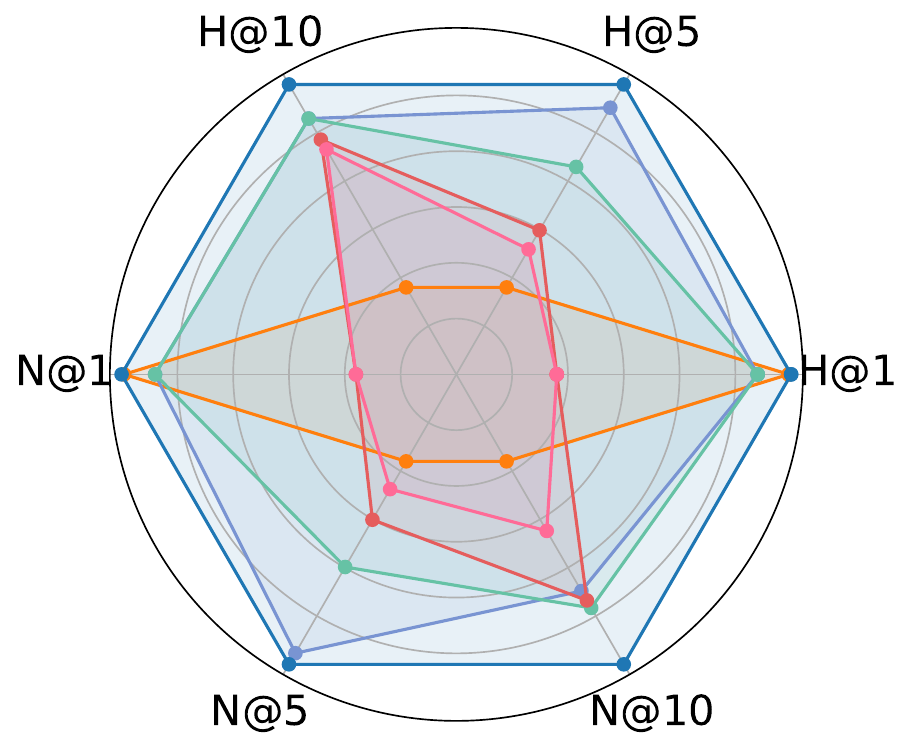}}
  \centerline{\footnotesize{(a) PHO}}
\end{minipage}\hspace{0.0in}
\begin{minipage}{0.3\linewidth}
  \centerline{\includegraphics[width=5cm]{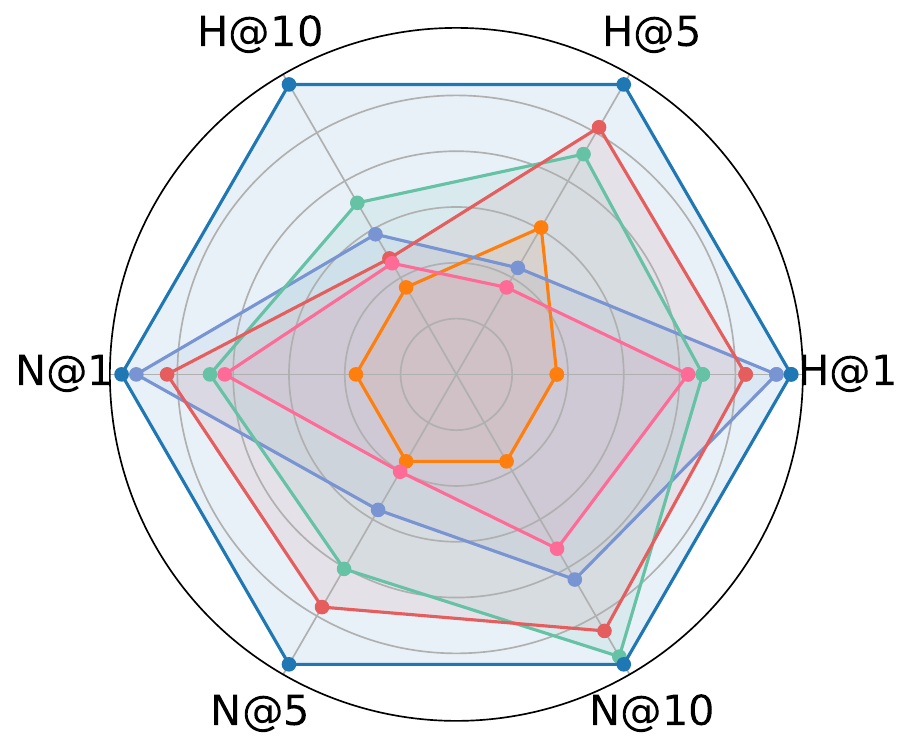}}
  \centerline{\footnotesize{(b) NY}}
\end{minipage}\hspace{0.25in}
\begin{minipage}{0.3\linewidth}
  \centerline{\includegraphics[width=7cm]{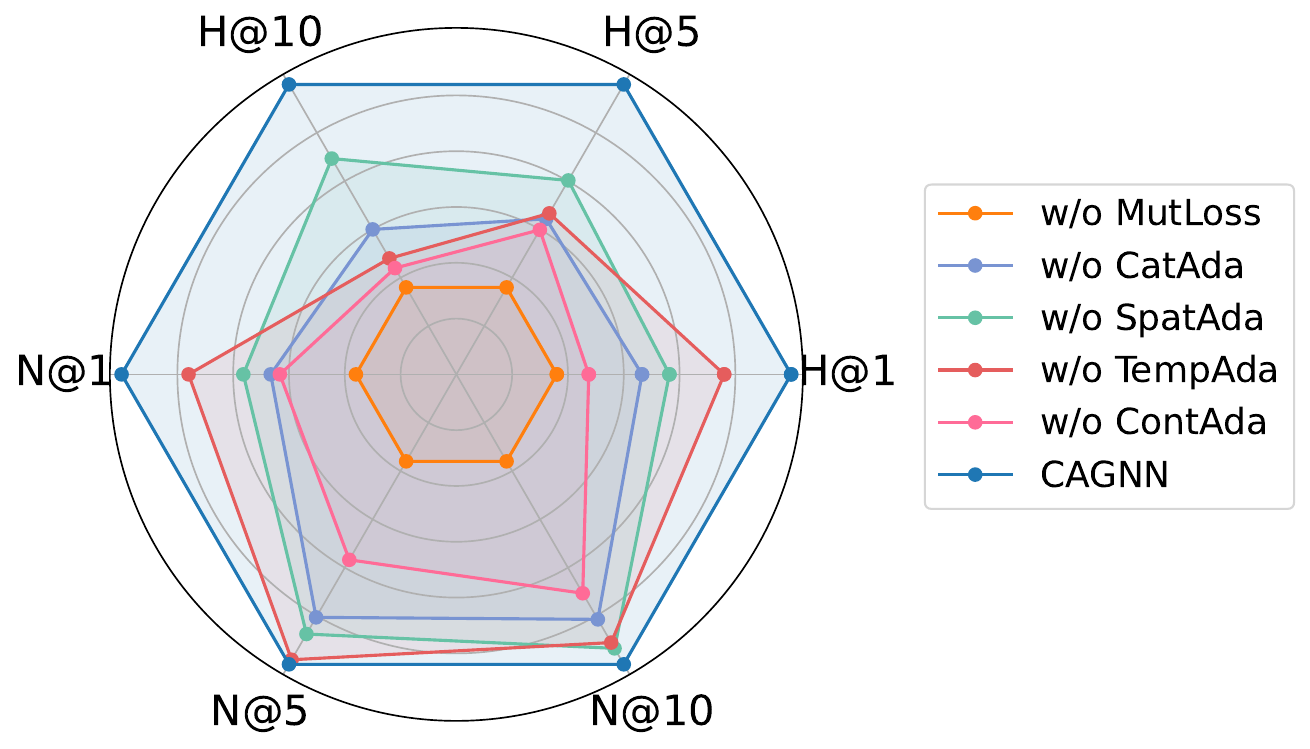}}
  \centerline{\footnotesize{(c) SIN}}
\end{minipage}
\caption{Performance comparison of different CAGNN variants on the three datasets in terms of HR (`H') and NDCG (`N').}
\label{fig:ablation}
\vspace{-0.05in}
\end{figure*}

\input{sensitive}

\subsection{Impacts of Different Components}
We conduct a comprehensive ablation study to evaluate the effectiveness of different components of CAGNN. As illustrated in Fig.~\ref{fig:ablation}, we compare the CAGNN model with five variants: 
(1) CAGNN w/o MutLoss, which removes the mutual enhancement loss $\mathcal{J}_m$ and retains only the prediction loss; 
(2) CAGNN w/o CatAda, which removes the Category-aware Adaptive Attention module, considering only spatial and temporal information; 
(3) CAGNN w/o SpatAda, which removes the Spatial-aware Adaptive Attention module, considering only category and temporal information; 
(4) CAGNN w/o TempAda, which removes the Temporal-aware Adaptive Attention module, considering only category and spatial information; 
and (5) CAGNN w/o ContAda, which removes all three adaptive attention modules and uses the standard GAT to model graph-based information.

Several key observations emerge from the results. \textbf{(1)} The CAGNN model consistently achieves the best performance across all datasets, implying the efficacy of each component.
\textbf{(2)} CAGNN w/o MutLoss shows the most significant performance drop, underscoring the importance of the mutual enhancement mechanism in integrating graph-based and sequential components, as well as the effectiveness of leveraging collaborative information from the graph structure.
\textbf{(3)} Removing any single context-aware adaptive attention component (CAGNN w/o CatAda, CAGNN w/o SpatAda, and CAGNN w/o TempAda) leads to notable performance degradation, demonstrating that each context (i.e., category, spatial, and temporal) contributes meaningfully to enhancing the graph attention mechanism.
\textbf{(4)} CAGNN w/o ContAda, which removes all context-aware adaptive attention modules, performs worse than the three variants that remove only one context. 
This suggests that jointly modeling multiple contextual factors is more effective than using any individual factor in isolation.
\textbf{(5)} The performance gap between CAGNN and CAGNN w/o ContAda confirms that the context-adaptive attention mechanism effectively captures collaborative and context-dependent patterns. As demonstrated in Section~\ref{sec:proof}, this enhances the expressiveness of POI representations and ultimately improves recommendation accuracy.

\subsection{Impacts of Important Hyper-parameters}
Fig.~\ref{fig:SIN_para} illustrates the impact of key hyperparameters on CAGNN, including Embedding size $dim$, the weight of mutual enhancement loss ($\beta$), the layer of GAT ($L$), the layer of Transformer. From these figures, several observations can be made.
\textbf{(1)} As $dim$ increases, both HR and NDCG show an upward trend, reaching peak performance around 120-150 dimensions. Beyond this range, further increases in $dim$ may lead to overfitting, causing performance to plateau or decline. 
\textbf{(2)} The weight of mutual enhancement loss ($\beta$) affects performance differently across datasets. For PHO and NY, performance peaks at $\beta = 0.7$, while for SIN, the optimal value is $\beta = 1$. However, for all datasets, performance first increases to a peak and then slightly drops with higher $\beta$ values. This indicates that excessive weight on mutual enhancement loss may impair the model’s accuracy.
\textbf{(3)} The number of GAT layers ($L$) impacts performance depending on the dataset characteristics. For PHO, optimal performance occurs at $L = 2-3$, while for NY, it peaks at $L = 5$. This difference may be due to varying data sparsity.
\textbf{(4)} For Transformer layers, optimal performance is observed at 2-3 layers for SIN and NY, with performance deteriorating at higher values. This suggests that deeper Transformer architectures are not beneficial for these datasets.
Similar observations are noted across all metrics with K = \{1, 5\}, but are omitted due to space constraints.

\begin{figure*}
    \centering
    \begin{subfigure}{0.3\textwidth}
        \centering
        \begin{tikzpicture}
            \begin{axis}[
                width=1\textwidth,
                height=0.65\textwidth,
                xlabel={Learned edge attention},
                ylabel={Log-scaled edge count },
                ymode=log,  
                ymin=1,
                xtick={0.0, 0.2, 0.4, 0.6, 0.8, 1.0}, 
            xticklabels={0.0, 0.2, 0.4, 0.6, 0.8, 1.0},
                ylabel style={yshift=-9pt, font = \small},
                tick label style={font = \footnotesize},
                legend style={at={(0.5,1.45)}, font=\scriptsize, anchor=north, legend columns=1, draw=none},
                legend image post style={scale=0.6}
            ]
        
                \addplot[smooth, thick, mark=square, mark size=1.5pt, orange] coordinates {
                    (0.0,404) (0.1,33) (0.2,36) (0.3,22) (0.4,14) (0.5,12)  (0.6,20) (0.7,18) (0.8,34) (0.9,5)
                };
                \addplot[smooth, thick, mark=o, mark size=1.5pt, black] coordinates {
                    (0.0,335) (0.1,53) (0.2,58) (0.3,37) (0.4,31) (0.5,30) (0.6,28) (0.7,10) (0.8,14) (0.9,2)
                };
                \legend{Context-adaptive attention mechanism, GAT}
        
            \end{axis}
        \end{tikzpicture}
        \vspace{-0.2in}
        \caption{\footnotesize{PHO}}
    \end{subfigure}
    \begin{subfigure}{0.3\textwidth}
        \centering
        \begin{tikzpicture}
            \begin{axis}[
                width=1\textwidth,
                height=0.65\textwidth,
                xlabel={Learned edge attention},
                ylabel={Log-scaled edge count },
                ymode=log,  
                ymin=10^2,
                xtick={0.0, 0.2, 0.4, 0.6, 0.8, 1.0}, 
            xticklabels={0.0, 0.2, 0.4, 0.6, 0.8, 1.0},
                ylabel style={yshift=-9pt, font = \small},
                tick label style={font = \footnotesize},
                legend style={at={(0.5,1.45)}, font=\scriptsize, anchor=north, legend columns=1, draw=none},
                legend image post style={scale=0.6}
            ]
        
                \addplot[smooth, thick, mark=square, mark size=1.5pt, orange] coordinates {
                    (0.0,11470) (0.1,968) (0.2,856) (0.3,691) (0.4,402) (0.5,426)  (0.6,567) (0.7,440) (0.8,544) (0.9,230)
                };
                \addplot[smooth, thick, mark=o, mark size=1.5pt, black] coordinates {
                    (0.0,10149) (0.1,1170) (0.2,1055) (0.3,968) (0.4,754) (0.5,902) (0.6,613) (0.7,294) (0.8,321) (0.9,168)
                };
                \legend{Context-adaptive attention mechanism, GAT}
        
            \end{axis}
        \end{tikzpicture}
        \vspace{-0.2in}
        \caption{\footnotesize{NY}}
    \end{subfigure}
    \begin{subfigure}{0.3\textwidth}
        \centering
        \begin{tikzpicture}
            \begin{axis}[
                width=1\textwidth,
                height=0.65\textwidth,
                xlabel={Learned edge attention},
                ylabel={Log-scaled edge count },
                ymode=log,  
                ymin=10^2,
                xtick={0.0, 0.2, 0.4, 0.6, 0.8, 1.0}, 
            xticklabels={0.0, 0.2, 0.4, 0.6, 0.8, 1.0},
                ylabel style={yshift=-9pt, font = \small},
                tick label style={font = \footnotesize},
                legend style={at={(0.5,1.45)}, font=\scriptsize, anchor=north, legend columns=1, draw=none},
                legend image post style={scale=0.6}
            ]
        
                \addplot[smooth, thick, mark=square, mark size=1.5pt, orange] coordinates {
                    (0.0,13445) (0.1,839) (0.2,637) (0.3,638) (0.4,398) (0.5,452)  (0.6,579) (0.7,436) (0.8,517) (0.9,147)
                };
                \addplot[smooth, thick, mark=o, mark size=1.5pt, black] coordinates {
                    (0.0,12879) (0.1,928) (0.2,899) (0.3,956) (0.4,573) (0.5,631) (0.6,398) (0.7,380) (0.8,434) (0.9,124)
                };
                \legend{Context-adaptive attention mechanism, GAT}
        
            \end{axis}
        \end{tikzpicture}
        \vspace{-0.2in}
        \caption{\footnotesize{SIN}}
    \end{subfigure}
    \vspace{-0.1in}
    \caption{Comparison of learned message weight distributions for different models on the three datasets.}
    \label{fig:attention_weights}
    \vspace{-0.1in}
\end{figure*}
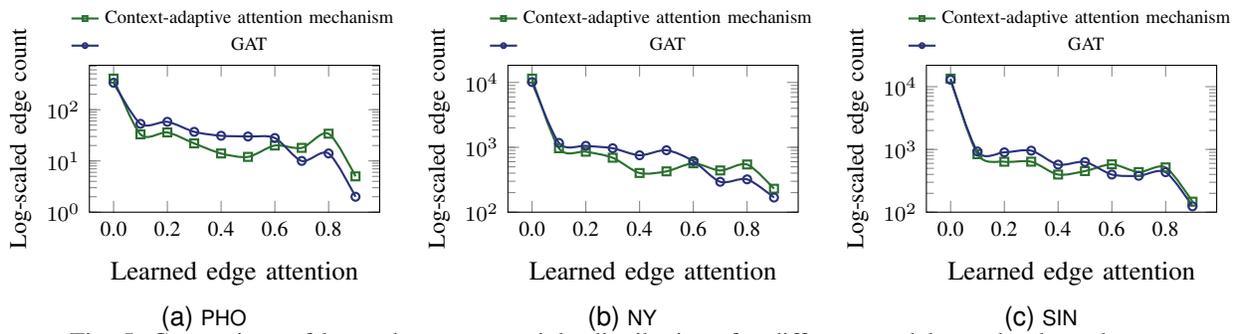
\begin{figure}[t]
\centering
\graphicspath{{}} 
\begin{minipage}{0.9\linewidth}
  \centerline{\includegraphics[width=\textwidth]{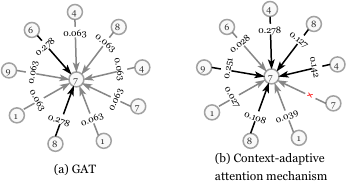}}
\end{minipage} 
\vspace{-0.05in}
\caption{Case study of attention learned by GAT and the context-adaptive attention mechanism for a randomly selected node.
Each circle represents a POI node, with the number inside indicating its category. Black edges denote neighbors assigned high attention weights (i.e., contextually relevant), while grey edges indicate lower-weight neighbors. Edges marked with a red ‘\textcolor{red}{x}’ are completely suppressed by the context-adaptive mechanism. Categories are defined as follows: 1: ‘Arts \& Entertainment',
 2: ‘College \& University',
 3: ‘Drink',
 4: ‘Food',
 5: ‘Nightlife Spot',
 6: ‘Outdoors \& Recreation',
 7: ‘Professional \& Other Places',
 8: ‘Residence',
 9: ‘Shop \& Service',
 10: ‘Travel \& Transport'. }
\label{fig:case}
\end{figure}

\subsection{Comparisons of Learned Graph Attention Weights}

In this section, we analyze the effect of the context-adaptive mechanism on the edge attention of the graph.
We compare the distribution of the edge attention learned by the context-adaptive attention mechanism and GAT. 
The corresponding results across all three datasets are shown in Fig.~\ref{fig:attention_weights}. 
A bin width of 0.1 is used, where, for example, 0.0 on the x-axis represents the attention range [0,0.1). The y-axis represents log-scaled edge counts. Since the original edge count and percentage of edges in each bin vary widely, we apply log-scaling to facilitate trend observation.
Two observations can be made based on the result. \textbf{(1)} Compared to GAT, the context-adaptive attention mechanism assigns zero weights to more edges, indicating its effectiveness in filtering out less relevant neighbors and improving propagation efficiency by focusing on more informative ones.
\textbf{(2)} The context-adaptive mechanism produces a more polarized attention distribution, with fewer edges assigned mid-range weights (0.4–0.6) and more edges falling into low (0–0.3) or high (0.7–1) ranges. This suggests improved discriminative ability in distinguishing relevant edges from less meaningful ones, rather than assigning ambiguous weights.

\subsection{Case Study}
To further assess the effectiveness of the proposed context-adaptive attention mechanism, we conduct a case study comparing its learned edge attention with that of standard GAT for a randomly selected central node from the NY dataset. As shown in Fig.~\ref{fig:case}, each circle represents a POI node, and the number inside indicates its category (see caption for details).
Several key observations can be drawn from the figure:
\textbf{(1)} The attention weights learned by GAT are relatively uniform across neighboring nodes, whereas the context-adaptive attention mechanism yields more diverse values, reflecting its ability to capture fine-grained contextual differences in transitions.
\textbf{(2)} The context-adaptive mechanism assigns lower attention weights to semantically less relevant neighbors (grey edges), with values such as 0.063 in Fig.\ref{fig:case}(a) and an average of approximately 0.03 in Fig.\ref{fig:case}(b).
\textbf{(3)} In some cases, the context-adaptive attention mechanism completely suppresses irrelevant transitions. This aligns with the distribution shown in Fig.~\ref{fig:attention_weights}, where more edges receive zero weights compared to GAT.
\textbf{(4)} The mechanism assigns higher weights to semantically relevant neighbors. Take the category context as an example, the central node (category 7: ‘Professional \& Other Places') receives higher attention weight from nodes in categories 4 (‘Food'), 8 (‘Residence'), and 9 (‘Shop \& Service'), which often co-occur with office locations in real-world mobility patterns.
These findings suggest that the context-adaptive attention mechanism captures collaborative and context-dependent transition patterns more effectively than standard GAT, leading to more expressive and context-aware POI representations.

\section{Conclusions and Future Work}

We identify two key limitations of existing GNN-based next POI recommendation models, including the lack of context-aware edge modeling and dominated sequential models. To address those limitations, we propose CAGNN, a novel framework that dynamically adjusts attention weights by jointly modeling multiple contextual factors and facilitates mutual enhancement between graph-based and sequential components.
The Graph-based User Preference Extractor employs a Context-Adaptive Attention mechanism to dynamically adjust GAT attention weights by jointly integrating category, spatial, and temporal contextual factors. This enhances the modeling of collaborative and context-dependent transitions.
The Sequential User Preference Extractor leverages a Transformer to model users’ short-term sequential preferences.
The Graph-Sequential Mutual Enhancement Module aligns POI embeddings from both extractors using KL divergence, allowing bidirectional knowledge transfer between the graph and sequential representations.
Theoretical analysis shows improved POI representation expressiveness, and experiments on three datasets confirm that CAGNN outperforms SOTAs.
These results highlight the effectiveness of integrating context-aware information into graph modeling. Future work may explore enhanced graph construction strategies and broader applications of graph attention mechanisms.

\section*{Acknowledgments}
This work was supported by the China Scholarship Council (Project ID: 202408130137). We used ChatGPT-3.5 to improve the manuscript’s readability; all content was reviewed and finalized by the authors.



\bibliographystyle{IEEEtran}
\bibliography{reference.bib}

\vspace{-0.3in}
\begin{wrapfigure}{l}{0.1\textwidth}
\vspace{-10pt}
\includegraphics[width=0.11\textwidth]{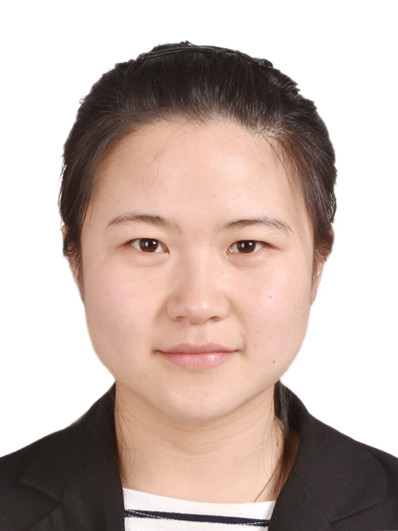}
\end{wrapfigure}
\begin{IEEEbiographynophoto}{Yu Lei} is currently pursuing the Ph.D. degree in the School of Information Science and Engineering, Yanshan University, China. And she is now a visiting student at Center for Frontier AI Research (CFAR), Institute of High Performance Computing (IHPC), Agency for Science, Technology and Research (A*STAR), Singapore, supported by the China Scholarship Council.
Her research topic includes Debiasing in recommendation, Graph Neural Networks and point-of-interest (POI) recommendation. 
\end{IEEEbiographynophoto}

\vspace{-0.3in}
\begin{wrapfigure}{l}{0.1\textwidth}
\vspace{-10pt}
\includegraphics[width=0.11\textwidth]{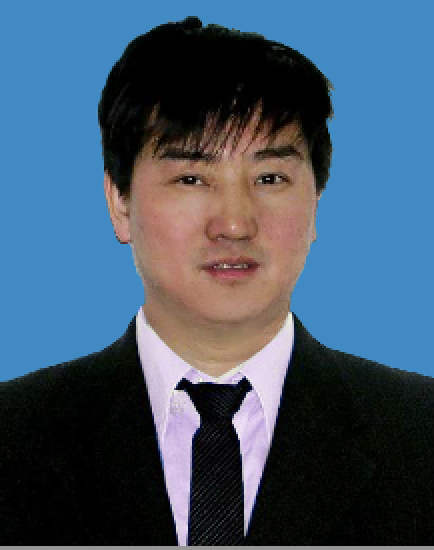}
\vspace{-25pt}
\end{wrapfigure}
\begin{IEEEbiographynophoto}{Limin Shen} (\textit{Member}, IEEE) received the B.S. degree from Yanshan University, China, and the M.S. degree from Hefei University of Technology in 1987. From 2005 to 2007, he was a visiting scholar at the Illinois Institute of Technology (IIT), USA. After returning to China in 2007, he served as the Associate Dean of the School of Information Science and Engineering at Yanshan University. He is currently a Professor and Ph.D. supervisor at Yanshan University. His research interests include recommender systems, data-driven security, service computing, and cooperative defense.
\end{IEEEbiographynophoto}

\vspace{-0.3in}
\begin{wrapfigure}{l}{0.1\textwidth}
\vspace{-10pt}
\includegraphics[width=0.11\textwidth]{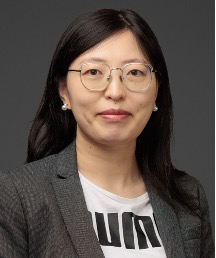}
\vspace{-20pt}
\end{wrapfigure}
\begin{IEEEbiographynophoto}{Zhu Sun} is currently an Assistant Professor at the Information Systems Technology and Design Pillar, Singapore University of Technology and Design. She received her Ph.D. degree from Nanyang Technological University (NTU), Singapore, in 2018. Her main research topic is artificial intelligence, specializing in trustworthy recommender systems. She has published papers in many leading conferences and journals, including SIGIR, SIGKDD, NeurIPS, IJCAI, AAAI, TheWebConf, CIKM, RecSys, TPAMI, TKDE, and TNNLS. She is the AE with Electronic Commerce Research and Applications (ECRA) and ACM Transactions on Recommender Systems (TORS), and PC/Senior PC Member for KDD, SIGIR, IJCAI, AAAI, SDM, CIKM, RecSys, etc. 
\end{IEEEbiographynophoto}

\vspace{-0.3in}
\begin{wrapfigure}{l}{0.1\textwidth}
\vspace{-10pt}
\includegraphics[width=0.11\textwidth]{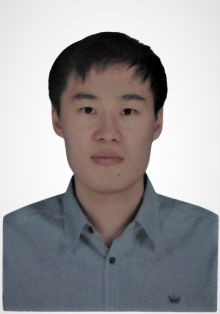}
\vspace{-20pt}
\end{wrapfigure}
\begin{IEEEbiographynophoto}{Tiantian He} (\textit{Member}, IEEE) received the Ph.D. degree in Computer Science from the Hong Kong Polytechnic University in 2017. Currently, He is a Senior Scientist at Center for Frontier AI Research (CFAR), Institute of High Performance Computing (IHPC), Singapore Institute of Manufacturing Technology (SIMTech), Agency for Science, Technology and Research (A*STAR). His research interests include artificial intelligence, machine learning, data-centric transfer optimization, and data mining. His work has been published in top conferences and journals, including NeurIPS, AAAI, ICDM, Artificial Intelligence, TKDE, TSC, and TCYB. He also serves as an AE for Memetic Computing, and PC member/Area Chair for NeurIPS, ICML, ICLR, AISTATS, IJCNN, etc.
\end{IEEEbiographynophoto}

\vspace{-0.3in}
\begin{wrapfigure}{l}{0.1\textwidth}
\vspace{-10pt}
\includegraphics[width=0.11\textwidth]{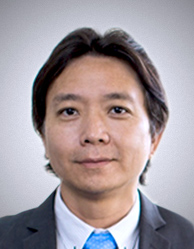}
\vspace{-25pt}
\end{wrapfigure}
\begin{IEEEbiographynophoto}{Yew-Soon Ong} (\textit{Fellow}, IEEE) received the Ph.D. degree in artificial intelligence in complex design from the University of Southampton, Southampton, U.K., in 2003. He is a President’s Chair Professor of Computer Science with Nanyang Technological University (NTU), Singapore, and holds the position of Chief Artificial Intelligence Scientist with the Agency for Science, Technology and Research (A*STAR), Singapore. At NTU, he serves as the Director of the Data Science and Artificial Intelligence Research and the Co-Director of the Singtel-NTU Cognitive and Artificial Intelligence Joint Laboratory. His research interest is in artificial and computational intelligence. Dr. Ong has received several IEEE outstanding paper awards and was listed as a Thomson Reuters Highly Cited Researcher and among the World’s Most Influential Scientific Minds. He was the founding EIC of IEEE Transactions on Emerging Topics in Computational Intelligence and AE of several others, including IEEE Transactions on Artificial Intelligence.
\end{IEEEbiographynophoto}


\end{document}